%Paper: hep-th/9210120
%From: adhar%theory%tifrvax.BITNET@CEARN.cern.ch
%Date: Thu, 22 Oct 92 12:42:03 IST
%Date (revised): Tue, 3 Nov 92 13:23:31 IST

%%%%%%%%%%%            PLEASE USE PHYZZX     %%%%%%%%%%%%%%%%%%%%
\input phyzzx.tex
\pubnum{92-63}
\date{October, 1992}
\titlepage
\title{\fourteenbf Stringy quantum effects in 2-dimensional Black-Hole}
\author{Avinash Dhar\foot{adhar@tifrvax.bitnet}, Gautam
Mandal\foot{mandal@tifrvax.bitnet} and Spenta R.
Wadia\foot{wadia@tifrvax.bitnet}}
\address{Tata Institute of
Fundamental Research, Homi Bhabha Road, Bombay 400 005, India}
\abstract{We discuss the classical 2-dim. black-hole in the framework of
the non-perturbative formulation (in terms of non-relativistic fermions) of
$c=1$
string field theory.  We identify an off-shell operator whose classical
equation of motion is that of tachyon in the classical graviton-dilaton
black-hole background.  The black-hole `singularity' is identified with
the fermi surface in the phase space of a single fermion, and as such is a
consequence of the semi-classical approximation.  An exact treatment
reveals that stringy quantum effects wash away the classical singularity.}

\endpage

\noindent {\bf 1. ~\underbar{Introduction}}

The semiclassical quantization of black hole solutions of classical
general relativity leads to a well-known paradox (which was first raised by
Hawking) of pure states evolving into mixed states.  Such a circumstance
occurs when a neutral black-hole evaporates completely due to Hawking
radiation.  There are various
viewpoints which either call for a modification of the present formulation
of quantum mechanics or present a resolution within the existing framework of
quantum mechanics.  However, it is fair to say that presently this issue
remains unresolved.  For recent reviews see [1-3].

One of the proposals to resolve Hawking's paradox is that we view
gravitation as naturally embeded in string theory.  Then one may be
tempted to argue that in the process of evaporation when the mass of the
black-hole is approximately planck mass, stringy fluctuations would have
to be included and these may smoothen out the classical curvature
singularity of the black-hole.  The problem with this proposal has been
that the known formulations of string theory in physical dimensions are
perturbative and as such are not equipped to address the question of
singularities.

However, it is very fortunate that 2-dimensional string theory has a
black-hole solution [4,5] and at the same time it has a non-perturbative
formulation in terms of non-interacting non-relativistic fermions
whose single particle hamiltonian is $h(p,q) = {1\over 2} (p^2-q^2)$ [6,7].
This fortunate circumstance can be exploited to answer some of the
issues of black-hole physics.

In this note we discuss the classical black-hole of the 2-dimensional
string theory in the non-perturbative framework of the fermion field
theory that is dictated by the $W_\infty$ symmetry algebra which is a one
parameter deformation of the algebra of
area preserving diffeomorphisms.  The deformation parameter is the string
coupling $g_{str} = \hbar \sim {1\over\mu}$.  The fundamental operator
of this formulation is $u(p,q,t) = \int dx \psi^+\left(q - {x \over
2}\right)~e^{-ipx} \psi \left(q + {x \over 2}\right)$, whose classical
limit gives the phase space distribution of fermions.  $u(p,q,t)$
transforms covariantly under $W_\infty$.  The string coupling,
among other things, labels the co-adjoint orbits of
$W_\infty$.  Motivated by the fact that the classical fermion trajectories
are hyperbolas in the $(p,q)$ phase plane we introduce a hyperbolic
measure on the phase plane
and define a transform of $u(p,q,t):~\phi(p,q,t) = \int {dp'~dq' \over
|p^{\prime 2}-q^{\prime 2}|^{1/2}} u(p' + p,q'+q,t)$.

The classical equation for $\phi (p,q,t)$ in the phase plane turns out
to be that of `massless' tachyon propagating in classical
black-hole background of the 2-dimensional string theory.  On-shell,
$\phi(p,q,t)$ is a function of only the two combinations $u = {1\over2}
(p+q)e^{-t}$ and $v = {1\over2} (p-q)e^t$.  The black-hole
singularity $uv = {\mu\over 2}$ precisely maps onto the hyperbola that
defines the
fermi surface in the phase plane, ${1\over2}(p^2 - q^2) = \mu$.  However,
in this non-perturbative formulation we can go beyond the semi-classical
calculation.  An exact computation of the tachyon background gives an
answer that is finite at $uv = \mu/2$.  From a detailed analysis we find
that near the black-hole singularity higher loop stringy effects become
important and in fact wipe out the singularity.  This provides
an explicit demonstration that stringy effects (at least in the lower
dimensional model) remove the curvature singularity.  The qualitative
picture of black-holes in the fermi fluid picture has been previously
discussed by us [8].

The plan of this paper is as follows.  In the next section we review some
relevant features of the non-perturbative formulation of $c=1$ string
field theory developed in [9].  In Sec. 3 we derive the Wheeler-deWitt
equation for the loop operator in the classical limit.  In Sec. 4 we
introduce the new tachyon operator and the mapping from the single fermion
phase space to
the $u,v$ plane.  Using this mapping and the Wheeler-deWitt equation we
then show that the tachyon satisfies equation (39) of [4].  We also
discuss the $W_\infty$ algebraic structure related to the tachyon
operator.  In Sec. 5 we present an exact calculation of the
one-point function of this operator in the fermion field theory.  This
one-point function is finite at the classical black-hole singularity.  We
discuss in detail how stringy loop corrections wipe out the singularity.
\bigskip

\noindent {\bf 2. \underbar{$c=1$ String Field Theory}:}

\nobreak
We briefly review the non-perturbative formulation of the $c=1$ string field
theory.  As is well-known this theory is exactly described by
non-relativistic fermions moving in a background hamiltonian.  The double
scaled field theory corresponds to the hamiltonian $h(p,q) = {1\over2}
(p^2 - q^2)$.  Since the fermion number is held fixed the basic
excitations are described by the bilocal operator $\phi(x,y,t) = \psi(x,t)
\psi^+(y,t)$ or equivalently its transform
$$
u(p,q,t) = \int^{+\infty}_{-\infty} dx~\psi^+\left(q - {x \over 2}\right)
e^{-ipx} \psi\left(q + {x \over 2}\right)
\eqno (1)
$$
The expectation value of this operator in a state is the fermion
distribution function in phase space.  Eqn. (1) also has the important
property that given a ``classical function'' $f(p,q,t)$ in the phase
space, we have an operator in the fermion field theory
$$
{\cal O}_f = \int dp~dq~f(p,q,t) u(p,q,t) = \int dx~\psi^+(x,t) \hat
f(\hat x,\hat p,t) \psi(x,t)
$$
where $\hat f(\hat x,\hat p)$ is the Weyl-ordered operator corresponding to
the classical function $f(p,q,t)$.  For example, vector fields
corresponding to the functions $f_{\alpha\beta} (p,q) =
e^{i(p\beta-q\alpha)}$ satisfy the classical algebra $\omega_\infty$ of
area-preserving diffeomorphisms.  The corresponding quantum operators in
the fermion field theory
$$
\tilde u(\alpha,\beta,t) = \int {dp~dq \over
(2\pi)^2} e^{i(p\beta-q\alpha)} u(p,q,t)
\eqno (2)
$$
satisfy the $W_\infty$ algebra (a one-parameter deformation of
$\omega_\infty$) [9]\foot{We have in our previous works also
used the notation $W(\alpha,\beta,t)$ for $\tilde u(\alpha,\beta,t)$.}
$$
[\tilde u(\alpha,\beta,t),\tilde u(\alpha',\beta',t)] = 2i \sin {\hbar
\over 2} (\alpha\beta' - \beta\alpha') \tilde u(\alpha + \alpha',\beta +
\beta',t)
\eqno (3)
$$

An exact boson representation of the fermion field theory that reflects
the $W_\infty$ symmetry can be acheived in terms of the 3-dim. field
$u(p,q,t)$, provided we impose the constraints that follow from its
microscopic definition
$$
\cos {\hbar \over 2} \left(\partial_q \partial_{p'} - \partial_{q'}
\partial_p\right) u(p,q,t) u(p',q',t)\bigg|_{{p' = p \atop q' = q}} = u
(p,q,t)
\eqno (4)
$$
$$
\int {dp~dq \over 2\pi \hbar} u(p,q,t) = N
\eqno (5)
$$
where $N$ is the total number of fermions.  Also the equation of motion
that follows from the definition (1) is
$$
(\partial_t + p\partial_q + q\partial_p) u(p,q,t) = 0
\eqno (6)
$$
The equation for $\tilde u(\alpha,\beta,t)$ is
$$
(\partial_t + \alpha\partial_\beta + \beta\partial_\alpha) \tilde
u(\alpha,\beta,t) = 0
\eqno (7)
$$
The constraints (4) and (5) in fact specify a co-adjoint orbit of
$W_\infty$, and the classical action is constructed using the
method of Kirillov
$$
\eqalign{
S[u,h] = &\int ds~dt \int {dp~dq \over 2\pi\hbar} u(p,q,t,s)
\hbar^2\left\{\partial_s u(p,q,t,s), \partial_t u(p,q,t,s)\right\}_{MB}
\cr & + \int dt \int {dp~dq \over 2\pi\hbar} u(p,q,t) h(p,q).}
\eqno (8)
$$
when $\{~,~\}_{MB}$ is the Moyal bracket (for details see [9]).

Let us now breifly indicate the classical limit of the string theory
$(\hbar \rightarrow 0)$.  In this limit the constraint (4) implies that
$u(p,q,t)$ is a characteristic function of a region of phase space
specified by a boundary [10-12].  For example the ground state corresponds
to the
static solution $u(p,q) = \theta(h(p,q)-\mu)$, $\mu \sim
-{1\over\hbar}$.  The massless excitation (tachyon) corresponds to a curve
that is a small deviation from the fermi surface ${1\over2}(p^2-q^2) =
\mu$.
\bigskip

\noindent {\bf 3. \underbar{Equation of motion for the density field}:}

\nobreak
In this section we shall derive a second order equation for the fourier
transform of the fermion density field
$$
\tilde u_0(\alpha,t) \equiv \tilde u(\alpha,0,t) = \int {dp~dq \over
(2\pi)^2} e^{-iq\alpha} u(p,q,t)
\eqno (9)
$$
Consider the Taylor expansion
$$
\tilde u(\alpha,\beta,t) = \sum^\infty_{n=0} {\beta^n \over n!} \tilde u_n
(\alpha,0,t), ~~~\tilde u_n = \partial^n_\beta \tilde
u(\alpha,\beta,t)|_{\beta=0}.
$$
Then equation (7) implies that
$$
\partial_t \tilde u_n (\alpha,t) + \alpha \tilde u_{n+1} (\alpha,t) + n
\partial_\alpha \tilde u_{n-1} (\alpha,t) = 0, ~~~n = 0,1,\cdots
\eqno (10)
$$
For $n=0$ and $n=1$ we have
$$
\partial_t \tilde u_0 + \alpha \tilde u_1 = 0, ~~~\partial_t \tilde u_1 +
\alpha \tilde u_2 + \partial_\alpha \tilde u_0 = 0
$$
Eliminating $\tilde u_1$ form these, and taking a derivative w.r.t. $\mu$,
the fermi energy, we get the equation
$$
\left(\partial^2_t - \alpha \partial_\alpha\right)\partial_\mu\tilde
u_0 (\alpha,t) = \alpha^2 \partial_\mu\tilde u_2(\alpha,t)
\eqno (11)
$$
Now using the definition $h(p,q) = {1\over2} (p^2-q^2)$, it is easy to see
that
$$
\partial_\mu\tilde u_2(p,t) = (\partial^2_\alpha - 2\mu)\partial_\mu\tilde
u_0(p,t) + \tilde\epsilon(\alpha,t)
$$
where
$$
\tilde\epsilon(\alpha,t) = \int {dp ~dq\over (2\pi)^2}
e^{-i\alpha q} (h(p,q) - \mu) \partial_\mu u(p,q,t)
\eqno (12)
$$
$\tilde\epsilon$ is essentially a measure of the deviation of the energy
from the fermi level $\mu$.  Using (12) in (11) we get the basic
operator equation for $\partial_\mu\tilde u_0 (p,t)$,
$$
\left(\partial^2_t - \left(\alpha \partial_\alpha\right)^2
+ 2\alpha^2 \mu\right) \partial_\mu\tilde u_0(\alpha,t) = -2\alpha^2
\tilde\epsilon (\alpha,t)
\eqno (13)
$$
Note that the static part of the operator on the l.h.s. is the
Wheeler-deWitt operator.

Before proceeding further let us make a simple application of the operator
eqn. (13) and derive the Wheeler-deWitt equation.  If we define
the time-independent ground state expectation value of $\tilde
u_0(\alpha,t)$ to be $\psi_0(\alpha,\mu)$, then we see that in the
classical limit (13) reduces to
$$
[(\alpha \partial \alpha)^2 - 2\alpha^2 \mu] \partial_\mu
\psi_0(\alpha,\mu) = 0
\eqno (14)
$$
which is the Wheeler-deWitt equation for the `loop operator' (for
imaginary loop lengths) [13].  In arriving at (14) we have used that the ground
state expectation value $\langle u(p,q,t)\rangle_0 =
\theta(h(p,q)-\mu)$ in the classical limit, so that
$\tilde\epsilon$ on the r.h.s. of (13) vanishes in this limit.

Let us now derive the classical equation for the operator $\tilde
u_0(\alpha,t)$ in a state that is a small deviation from the classical
ground state.  As is well-known, such a state is described by a
characteristic function of phase space whose boundary is a small
area-preserving deviation from the fermi surface ${1\over2} (p^2 - q^2) =
\mu$ [10-12].  Call the region bounded by the hyperbola ${1\over2} (p^2 - q^2)
= \mu, ~q < 0$, $R_0$, and
the deformed region $R$.  By small deformation
we mean $|{\delta E \over \mu}|
\equiv |{h(p,q) - \mu \over \mu}| \ll 1$, for $(p,q) \in
\delta R \equiv R-R_0$.
(Note that $\int_{\delta R} \delta E/\mu$ is in fact the
expansion parameter of the string perturbation theory).

Let us consider the expectation value of $\tilde u_0(\alpha,t)$ in the
state $|R \rangle$.  Then from (13) we have
$$
\left(\partial^2_t - (\alpha\partial_\alpha)^2 + 2\alpha^2\mu\right)
\partial_\mu\langle\tilde u_0(\alpha,t)\rangle_R = -2\alpha^2
\langle\tilde\epsilon (\alpha,t)\rangle_R
\eqno (15)
$$
Now
$$
\langle\tilde\epsilon(\alpha,t)\rangle_R = \int\int {dp~dq \over (2\pi)^2}
e^{-i\alpha q} \left(h(p,q) - \mu\right)
\partial_\mu\chi_{R(t)} (p,q)
\eqno (16)
$$
where $\chi_{R(t)} (p,q) = \langle u(p,q,t)\rangle_R$ is the
characteristic function of the region $R$.  Denoting $\chi_{R(t)} -
\chi_{R_0}$ by $\delta\chi(p,q,t)$, we have
$$
\eqalign{
\langle\tilde\epsilon(\alpha,t)\rangle_R = &\int {dp~dq \over
(2\pi)^2} e^{-i\alpha q} \left(h(p,q) - \mu\right)
\partial_\mu \chi_{R_0} (p,q) \cr &
+ \int {dp~dq \over (2\pi)^2} e^{-i\alpha q} \left(h(p,q) - \mu\right)
\partial_\mu \delta\chi(p,q,t)}
\eqno (17)
$$
Since $\chi_{R_0} (p,q) = \theta(h(p,q)-\mu)$, the first term
in the above vanishes.  Also
since $\left(h(p,q) - \mu\right)\allowbreak/\mu \ll 1$, for $\delta
R(t)$ a small region around the fermi surface, the second
term can be neglected compared to $\mu$.  The neglected terms
represent higher order terms in string perturbation theory.  Hence we have
$$
\left(\partial^2_t - (\alpha\partial_\alpha)^2 + 2\alpha^2\mu\right)
\partial_\mu \langle\tilde u_0(\alpha,t)\rangle_R = ({\rm
higher~orders~in~string~perturbation}).
\eqno (18)
$$
Defining $\psi(\alpha,t) = \langle\tilde u_0(\alpha,t)\rangle_R -
\psi_0(\alpha,\mu)$, where $\psi_0(\alpha,\mu)$ is a solution of the classical
Wheeler-deWitt equation (14), we get the basic classical equation for
$\psi(\alpha,t)$
$$
\left(\partial^2_t - (\alpha \partial_\alpha)^2 + 2\alpha^2 \mu\right)
\partial_\mu\psi(\alpha,t) = 0 + ``{\rm corrections}''
\eqno (19)
$$
The r.h.s. of the above equation is of course zero upto higher order
string loop ``corrections'', which can in principle be calculated from the
fermion theory.
\bigskip

\noindent {\bf 4. \underbar{\bf The hyperbolic transform}}

\nobreak
So far we have discussed the phase space density field $u(p,q,t)$ and the
generators $\tilde u(\alpha,\beta,t)$ of $W_\infty$ algebra.  We
now introduce
another field, $\phi(p,q,t)$, through a ``hyperbolic'' transform of
$u(p,q,t)$:
$$
\phi(p,q,t) \equiv \int {dp'~dq' \over |p^{\prime 2} - q^{\prime
2}|^{1/2}} u(p'+p,q'+q,t)
\eqno (20)
$$
The choice of the measure in (20) is motivated by the fact that the
classical trajectories of the fermions are hyperbolas in phase space.
Also note that the transform is defined off-shell.
Using the equation of motion (6) of $u(p,q,t)$ it is easy to see that
$\phi(p,q,t)$ satisfies an identical equation:
$$
(\partial_t + p\partial_q + q\partial_p) \phi(p,q,t) = 0
\eqno (21)
$$
In other words, on-shell, $\phi(p,q,t)$ is a function of only the two
combinations\foot{We hope there is no confusion
between the notation $u$ used below and that for the phase space density
$u(p,q,t)$.  The reason for using this notation is to facilitate contact
with earlier work.}
$$
u \equiv {1\over2} (p+q) e^{-t}, ~~~v \equiv {1\over2} (p-q) e^t.
\eqno (22)
$$
We will denote this function suggestively by $T(u,v)$, i.e.
$$
\phi(p,q,t) {\buildrel {\rm on-shell} \over \equiv} T(u,v)
\eqno (23)
$$
The remarkable point is that the function $T(u,v)$ satisfies, in the
classical
approximation, the equation of motion of tachyon in the background of
black-hole metric and dilaton fields in the conformal gauge, the
variables $u$ and $v$ being identified with the conformal coordinates of
2-dimensional string theory [4].

It is not very difficult to prove the above.  The first step is to recast
the r.h.s. of (20), on-shell (i.e. using the equation of motion of
$u(p,q,t))$, as explicitly a function of $u$ and $v$ only.  To do so one
needs to make the change of variables
$$
p' = E \cosh \theta, ~~~q' = E \sinh \theta, ~~~{\rm for} ~ p^{\prime 2} -
q^{\prime 2} \geq 0
$$
and \hfill (24)
$$
p' = E \sinh \theta, ~~~ q' = E \cosh \theta, ~~~{\rm for} ~ p^{\prime 2}
- q^{\prime 2} \leq 0
$$
in the integral.  Then, after some manipulations one can show that
$$
T(u,v) = \int^{+\infty}_{-\infty} dE \int^{+\infty}_{-\infty}
{}~d\theta\left[u(E,ue^\theta - ve^{-\theta},\theta) +
u(ue^\theta + ve^{-\theta}, E,\theta)\right]
\eqno (25)
$$
This provides an explicit demonstration of the fact that $\phi(p,q,t)$
satisfies the equation (21).  An operator similar to the first term in
(25) has previously been interpreted as tachyon in black-hole background
in [14].  We, however, believe that our symmetrical form is the more
natural one to use.  It is also important to realize that the form (25)
for our proposed tachyon operator has been derived from the off-shell
definition (20) after using the equation of motion for the $u$ field.

The second step is to recast (25) in terms of the field $\tilde
u(\alpha,\beta,t)$ by using
$$
u(p,q,t) = \int d\alpha~d\beta ~e^{i(q\alpha - p\beta)} \tilde
u(\alpha,\beta,t)
\eqno (26)
$$
This gives
$$
T(u,v) = 2\pi \int^{+\infty}_{-\infty} d\theta \int^{+\infty}_{-\infty}
d\alpha\left[e^{i\alpha(ue^\theta -
ve^{-\theta})} \tilde u(\alpha,0,\theta) + e^{-i\alpha(ue^\theta +
ve^{-\theta})} \tilde u(0,\alpha,\theta)\right]
\eqno (27)
$$

In the final step, let us define $\eta(u,v) \equiv T(u,v) - T_0(u,v)$,
where $T_0$
is the vacuum value of $T$ obtained from (27) by substituting the vacuum
value of $\tilde u(\alpha,\beta,\theta)$ in it.  Then, shifting $\tilde
u(\alpha,\beta,\theta)$ by its vacuum value and using the equations (14)
and (19) one can show that the fluctuations $\eta(u,v)$ satisfy the
equation
$$
\left[4\left(uv - {\mu \over 2}\right) \partial_u \partial_v
+ 2(u\partial_u + v\partial_v) +
1\right] \partial_\mu\eta(u,v) = 0 + {\rm ``corrections''}
\eqno (28)
$$
This is identical to equation (38) of ref. [4] in the conformal gauge.
The metric and dilaton background that enter in (28) are those of the
classical 2-dim. black-hole in the conformal gauge of the 2-dim. gravity.
It should be remarked that the full Kruskal extension emerges because we
are working with the entire $(p,q)$ phase plane of the fermion problem.

This identification of the tachyon field in the background of black-hole
metric and dilaton system with the (matrix model) fermion field theory
operator $\phi(p,q,t)$ defined in (20) is quite remarkable.  What is more
remarkable, however, is that (28) has solution which is singular at $uv
= {\mu \over 2}$, while equation (25), from which (28) was derived using
``reasonable'' assumptions, apparently has nothing singular about it at
$uv = {\mu \over 2}$.  What does $uv = {\mu \over 2}$ correspond to on the
phase plane?  The mapping we have constructed from the phase plane (at
each instant
of time) to the $(u,v)$ space given by (22) enables us to answer this
question easily.  The $uv = {\mu \over 2}$ singularity corresponds to the
classical single particle trajectory with energy $h(p,q) = {1\over2} (p^2
- q^2) = \mu$.  This identifies the classical black-hole mass with $\mu/2$
where $\mu$ is the
fermi level.  We mention that in [15], using continuum methods, the
classical black-hole mass was identified with the scale of the Liouville
theory (which is essentially $\mu$).
We thus arrive at the result that the fermi surface
of the classical fermi fluid theory gets mapped on to the
black-hole singularity in the $(u,v)$ plane.  Since there is nothing
singular about the fermi surface, it would seem that the singularity at $uv =
{\mu \over 2}$ in (28) is a result of neglecting the ``correction'' terms
in the classical approximation.  To resolve this issue one needs to
do an exact calculation.  Such a calculation can be done for the tachyon
background $T_0 (u,v)$.  Before turning to it, we close this section with a
discussion of the algebra satisfied by the operator $\phi(p,q,t)$.

Analogous to the generators $\tilde u(\alpha,\beta,t)$ of
$W_\infty$ algebra corresponding to $u(p,q,t)$, let us introduce
the operators $\tilde\phi(\alpha,\beta,t)$ corresponding to $\phi(p,q,t)$:
$$
\tilde\phi(\alpha,\beta,t) \equiv \int {dp~dq \over (2\pi)^2} e^{i(p\beta
- q\alpha)} \phi(p,q,t)
\eqno (29)
$$
Using (20), we get
$$
\tilde\phi (\alpha,\beta,t) = \left(\int {dp' ~dq' \over |p^{\prime 2} -
q^{\prime 2}|^{1/2}} e^{i(q'\alpha - p'\beta)}\right) \tilde
u(\alpha,\beta,t)
$$
The integral can be easily done using the hyperbolic parametrization given
in (24).  The result is
$$
\tilde\phi(\alpha,\beta,t) = {2 \over |\alpha^2 - \beta^2|^{1/2}} \tilde
u(\alpha,\beta,t)
\eqno (30)
$$
This simple proportionality of $\tilde \phi$ and $\tilde u$ leads to the
remarkable result that $\tilde \phi$ also satisfies a $W_\infty$
algebra:
$$
\left[\tilde\phi(\alpha,\beta,t),\tilde\phi(\alpha',\beta',t)\right] = 4i
\sin {\hbar \over 2} (\alpha\beta' - \alpha'\beta) {|(\alpha + \alpha')^2
- (\beta + \beta')^2|^{1/2} \over |\alpha^2 - \beta^2|^{1/2}
|\alpha^{\prime 2} - \beta^{\prime 2}|^{1/2}} \tilde\phi (\alpha +
\alpha',\beta + \beta',t)
\eqno (31)
$$
The structure constants are now, however, very different and in fact
singular.  But this singularity is merely a reflection of the singular
factor relating $\tilde u$ to $\tilde \phi$ in (30).  That there is a
$\omega_\infty$ algebra underlying the black-hole geometry has been
recently discussed in [16] in the context of $SL(2,R)/U(1)$ coset model
of the black-hole.  Here we see that this is true for the exact quantum
problem (except that the $\omega_\infty$ algebra gets modified to its
quantum counterpart $W_\infty$).
\bigskip

\noindent {\bf 5. \underbar{\bf Exact evaluation of the tachyon
background}}

\nobreak
The tachyon background $T_0 (u,v)$ is just the one-point function of the
operator given in (28) in the fermi vacuum.  The one-point function of the
operator $\tilde u(\alpha,\beta,\theta)$ in the fermi vacuum is a function
of $(\alpha^2 - \beta^2)$ only and can be shown to satisfy an exact
quantum generalization of the Wheeler-deWitt equation (14).  Because of
the dependence through $(\alpha^2 - \beta^2)$ only, it is enough to write
down the equation for $\langle \tilde u(\alpha,o,\theta)\rangle_0 \equiv
\psi(\alpha,\mu)$:
$$
[(\alpha \partial_\alpha)^2 - 2\mu\alpha^2 + \alpha^4/4] \partial_\mu
\psi(\alpha,\mu) = 0
\eqno (32)
$$
The other function $\psi'(\beta,\mu) \equiv \langle\tilde
u(0,\beta,\theta)\rangle_0$ can then be obtained from $\psi(\alpha,\mu)$
by the formal substitution $\alpha \rightarrow i\beta$.  Equation (32) was
first obtained in [13].  It can be shown to be a direct consequence of
the fermion equation of motion [17].

Using equation (32) and a similar one for $\partial_\mu \psi'(\beta,\mu)$
in (28) one can easily derive the following exact equation for the tachyon
background:
$$
\left[{1\over4} \partial^2_u \partial^2_v + 4\left(uv - {\mu \over
2}\right) \partial_u \partial_v + 2(u\partial_u + v\partial_v) +
1\right]\partial_\mu T_0 (u,v) = 0
\eqno (33)
$$
Since $T_0(u,v)$ is a function of the variable $\xi = uv$ only (as can be
easily deduced from (28) using that $\langle\tilde
u(\alpha,\beta,\theta)\rangle_0$ is independent of $\theta$), (33) can be
simplified to read
$$
\left[{1\over4} (\partial_\xi \xi \partial_\xi)^2 + 4(\xi - \mu/2)
\partial_\xi \xi \partial_\xi + 4\xi\partial_\xi + 1\right] T_0(\xi) = 0
\eqno (34)
$$
where we have defined $\partial_\mu T_0(u,v) \equiv T_0(\xi)$.

Evidently eqn. (34) does not show any singular behaviour for $\xi
\rightarrow \mu/2$.  One can reduce this equation to a hypergeometric
equation in momentum space (conjugate to $\xi$) and thus solve it.  We
will not do so here, but instead we will directly substitute the
expression for $\psi(\alpha,\mu)$ obtained from a computation in the
fermion field theory in (27) and evaluate $T_0(\xi)$.
It is important to realize that this obviates the need to discuss boundary
conditions in solving (34), since the fermion field theory gives a
definite answer for $\psi(\alpha,\mu)$.  An integral representation for
$\partial_\mu \psi(\alpha,\mu)$ has been given in [18]\foot{The prefactor
given in front of the integral in this reference is incorrect.  The
correct one has been given below.  Also, note the different sign
convention for the fermi level $\mu$.}.  It is
$$
\partial_\mu \psi(\alpha,\mu) = {1\over2\pi}~{\rm Re} \int^\infty_0 d\lambda
{e^{-i\mu\lambda + {i\over4} \alpha^2 \coth \lambda/2} \over \sinh
\lambda/2}
\eqno (35)
$$
The integral in (35) is defined for complex
$\alpha^2$ with a small positive imaginary part.  The result is then
analytically continued to real $\alpha^2$.  Similarly,
$$
\partial_\mu \psi'(\beta,\mu) = {1\over2\pi} ~{\rm Re} \int^\infty_0 d\lambda
{e^{i\mu\lambda + {i\over4} \beta^2 \coth {\lambda \over 2}} \over \sinh
\lambda/2}
\eqno (36)
$$
Using (35) and (36) in (27), we get after some manipulations
$$
\eqalign{
T_0(\xi) = (2\pi)^{3/2} ~{\rm Im} \int^\infty_0 d\lambda {e^{i\mu\lambda -
2i\xi \tanh {\lambda \over 2}} \over \sqrt{\sinh \lambda}}
\Bigg[&-e^{-i{\pi \over 4}} H^{(1)}_0 (2 |\xi| \tanh {\lambda \over 2})
\cr & + e^{i{\pi \over 4}} H^{(2)}_0 (2|\xi|\tanh {\lambda \over 2})\Bigg]}
\eqno (37)
$$
where $H^{(1)}_0 (z)$ and $H^{(2)}_0 (z)$ are standard combinations of
Bessel function [19].

Equation (37) reveals the source of the $uv = \xi \rightarrow \mu/2$
singularity in (28).  The classical limit in which (28) is valid
corresponds to restricting the range of $\lambda$-integration in (37) to
the region ${1 \over |\mu|} \ll \lambda \ll 1$, $|\xi|\lambda \gg 1$.  In
this range of
$\lambda$ we can approximately replace $2\tanh {\lambda \over 2}$ and
$\sinh \lambda$ by $\lambda$ and use the asymptotic expansions of $H^{(1)}_0$
and $H^{(2)}_0$ [20],
$$
\eqalign{
& H^{(1)}_0 (z) \sim \sqrt{2 \over \pi z} e^{i(z - \pi/4)} \cr &
H^{(2)}_0 (z) \sim \sqrt{2 \over \pi z} e^{-i(z-\pi/4)}}
$$
to get
$$
T_0 (\xi) \sim {\rm Re} \int^1_{1/\mu} {d\lambda \over \lambda} e^{i(\mu -
\xi)\lambda} (e^{-i|\xi|\lambda} + e^{i|\xi|\lambda})
\eqno (38)
$$
which gives the singular solution
$$
T_0 (\xi) \sim \ell n\left(\xi - {\mu \over 2}\right)
\eqno (39)
$$
obtained in [4,5].  The singularity in (38) at $\xi = \mu/2$ is, however,
clearly a result of the approximations made in deriving it since the full
expression in (37) is finite at $\xi = \mu/2$.  At this value of $\xi$
there is a cancellation of the terms in the exponent linear in $\lambda$
in (37), so one must retain higher order terms in $\lambda$ in the
expansion of $\tanh \lambda/2$.  The correct result is
$$
T_0 (\xi = \mu/2) \sim ~{\rm Re}~ \int^1_{1/\mu} {d\lambda \over \lambda}
e^{i {\mu \over 12} \lambda^3}
\eqno (40)
$$
which is obviously finite.  In fact, for $(\xi - \mu/2)$ very small,
$T_0 \sim a + b (\xi - \mu/2) +$ higher orders, where $a$ and $b$ are
constants.  Since the genus expansion
of $T_0(\xi)$ is
obtained by taking $\lambda \rightarrow \kappa \lambda$ in (37) and
expanding in powers of $\kappa$, we see that stringy higher loop effects
are responsible for the finiteness of $T_0 (\xi)$ at $\xi = \mu/2$.  We
wish to emphasize that this has been possible because the exact answer
(37) already sums string perturbation theory to all orders.  From
point of view of the exact equation (33) the higher derivative term is
important close to the classical singularity.  One might wonder about possible
geometrical interpretation of the exact equation (33).  Is there a
description at all in the neighbourhood of the classical singularity in
terms of backgrounds corresponding to graviton, dilaton and the other
higher modes of the string?  Or is
there a topological description without space time geometry?

In conclusion, we wish to point out that the identification of the tachyon
operator, (20), in fermion field theory corresponding to the classical
black-hole geometry in 2-dim. string theory, coupled with the fact that
the fermion theory can be treated exactly, has opened up the possibility
of understanding precisely how a consistent theory of quantum gravity
washes away the classical black-hole singularity.  It is to be hoped that
a similar mechanism is operative in more realistic higher dimensional
theories.

\noindent {\bf Acknowledgement:}  We would like to acknowledge A.M.
Sengupta for many discussions on black-hole physics.  One of us (SRW)
would also like to thank S.R. Das for useful discussions.

\noindent {\bf Note Added:}

\noindent(i)
The second-order differential equation for the tachyon background in
the classical approximation ((28) with $\eta(u,v)$ replaced by $T_0(u,v)$)
has two independent solutions one of which is regular at $uv= \mu/2$ and
the other singular. That the appropriate boundary condition is the one
which picks up the singular solution can be seen by directly evaluating
(20) for the fermi vacuum. In the classical limit, $\langle u(p,q,t)
\rangle_0 = \theta (h(p,q) - \mu)$. So we have, in the classical limit,
$$ \partial_\mu T_0(u,v) = -\int dp'dq'\, |(p-p')^2-(q-q')^2|^{-1/2}
\delta( {1\over 2}(p^{\prime 2} - q^{\prime 2}) - \mu)
\eqno (41)$$
Since $\mu$ is negative in our convention, we can use the change of variable
$$p'= E\sinh \theta, \qquad q'= E \cosh \theta $$
to do the above integration. The $E$-integration can be readily done using
the $\delta$-function. The result is
$$\eqalign{
\partial_\mu T_0(u,v) = -\int_{-\infty}^{\infty} d\theta \, & [
| 4uv + 2\mu + 2\sqrt{-2\mu}(u e^{-\theta} - v e^{\theta}) |^{-1/2}\cr
\, &+
| 4uv + 2\mu - 2\sqrt{-2\mu}(u e^{-\theta} - v e^{\theta}) |^{-1/2}
] \cr}
\eqno (42)
$$
For $uv\le 0$, we get
$$\eqalign{
\partial_\mu T_0(u,v) = -\int d\theta \, [
&| 4uv + 2\mu + 4\sqrt{-2\mu} | uv |^{1/2} \cosh\theta |^{-1/2}\cr
\,&+
| 4uv + 2\mu - 4\sqrt{-2\mu} | uv |^{1/2} \cosh\theta |^{-1/2}
] \cr}
\eqno (43)
$$
It is easy to see that the $\theta$-integration diverges logarithmically
for $uv \to \mu/2$,
$$ \partial_\mu T_0(u,v) \sim \ln (uv-\mu/2).$$
For $uv\ge 0$,
$$\eqalign{
\partial_\mu T_0(u,v) = -\int d\theta [
&| 4uv + 2\mu + 4\sqrt{-2\mu} | uv |^{1/2} \sinh\theta |^{-1/2}\cr
\,&+
| 4uv + 2\mu - 4\sqrt{-2\mu} | uv |^{1/2} \sinh\theta |^{-1/2}
] \cr}
\eqno (44)
$$
This is obviously non-singular. We see from this calculation, directly
from the classical fermi fluid theory, that the appropriate boundary
conditions on the differential equation for the tachyon in the black
hole metric is the one that picks out the singular solution.

\noindent (ii)
We thank E. Martinec  for pointing out to us that a tranform similar to
the first term in (25) has appeared in the context of the connection
between Liouville theory and the $SL(2,R)/U(1)$ black hole theory in a
note added in the published version of [15].

\endpage

\centerline{\underbar{\bf References}}

\item{[1]} E. Witten, IAS preprint, IASSNS-HEP-92/24.

\item{[2]} J.A. Harvey and A. Strominger, Chicago preprint, EFI-92-41,
hep-th/9209055.

\item{[3]} J. Preskill, Caltech preprint, CALT-68-1819, hep-th/9209058.

\item{[4]} G. Mandal, A.M. Sengupta and S.R. Wadia, Mod. Phys. Lett. A6
(1991) 1685.

\item{[5]} E. Witten, Phys. Rev. D44 (1991) 314.

\item{[6]} E. Brezin, C. Itzyksen, G. Parisi and J.B. Zuber, Comm. Math.
Phys. 59 (1978) 35.

\item{[7]} E. Brezin, V.A. Kazakov and Al.B. Zamolodchikov, Nucl. Phys.
338 (1990) 673; D.J. Gross and N. Milikovic, Nucl. Phys. B238 (1990) 217;
G. Parisi, Europhys Lett. 11 (1990) 595; P. Ginsparg and J. Zinn-Justin,
Phys. Lett. 240B (1990) 333; S.R. Das, A. Dhar, A. Sengupta and S.R.
Wadia, Mod. Phys. Lett. A5 (1990) 891.

\item{[8]} S.R. Wadia, Colloquium delivered at Tata Institute, Sept. 1992.

\item{[9]} A. Dhar, G. Mandal and S.R. Wadia, TIFR preprint,
TIFR-TH-92/40, (To appear in Mod. Phys. Lett. A).

\item{[10]}
A.M. Sengupta and S.R. Wadia, Int. J. Mod. Phys. A6 (1991) 1961;
A. Jevicki and
B. Sakita, Nucl. Phys. B165 (1980) 511; S.R. Das and A.
Jevicki, Mod. Phys. Lett. A5 (1990) 1639;
J. Polchinski, Nucl. Phys. B346 (1990) 253.

\item{[11]} S.R. Das, A. Dhar, G. Mandal and S.R. Wadia, Int. J. Mod. Phys.
A7 (1992) 5165.

\item{[12]} A. Dhar, G. Mandal and S.R. Wadia, TIFR preprint,
TIFR/TH/91-61 (To appear in Mod. Phys. Lett. A).

\item{[13]} G. Moore, Nucl. Phys. B368 (1992) 557.

\item{[14]} S.R. Das, TIFR preprint, TIFR/TH/92-62.

\item{[15]} E. Martinec and S. Shatashville, Nucl. Phys. B368 (1992) 338.

\item{[16]} T. Eguchi, H. Kanno and S.K. Yang, ``$W_\infty$ Algebra in
Two-Dimensional Black Hole'', Newton Institute Preprint NT-92004 (1992).
See also, S. Chaudhuri and J. Lykken, Fermilab Preprint 169/92 (1992).

\item{[17]} A. Dhar, G. Mandal and S.R. Wadia, unpublished; A. Dhar,
Lectures in the Spring School on Superstrings, Trieste, 1992.  To appear
in the proceedings.

\item{[18]} S.R. Das, A. Dhar, G. Mandal and S.R. Wadia, Mod. Phys. Lett.
A7 (1992) 937.

\item{[19]} I.S. Gradshteyn and I.M. Ryzhik, Table of Integrals, Series
and Products (Academic Press, New York) p. 955.

\item{[20]} I.S. Gradshteyn and I.M. Ryzhik, Table of Integrals, Series
and Products
(Academic Press, New York) p. 962 items 3 and 4.

\end